\documentclass[12pt]{article}

\usepackage{amssymb}
\usepackage{amsmath}
\usepackage{graphicx}

\newcommand{\eq}{\begin{equation}}
\newcommand{\eqx}{\end{equation}}
\newcommand{\eqs}{\begin{equation*}}
\newcommand{\eqsx}{\end{equation*}}
\newcommand{\eqn}{\begin{eqnarray}}
\newcommand{\eqnx}{\end{eqnarray}}
\newcommand{\alg}{\begin{align}}
\newcommand{\algx}{\end{align}}

\newcommand{\f}[2]{\frac{#1}{#2}}

\newcommand{\lm}{\lambda}

\newcommand{\dl}{\delta}

\newcommand{\eps}{\varepsilon}
\newcommand{\qqqq}{\quad\quad\quad\quad}

\newcommand{\slii}{{\tt sl(2)}\ }

\DeclareMathOperator{\arcsinh}{arcsinh}

\newcommand{\tr}{\mbox{\rm tr}\,}
\newcommand{\res}{\mbox{\rm res}\,}
\newcommand{\nn}{{\cal N}}
\newcommand{\oo}[1]{{\cal O}\left(#1\right)}

\newcommand{\sca}{S_{0}}
\newcommand{\mat}{S_{\boxplus}}
\newcommand{\dress}{S_{\sigma}}
\newcommand{\ex}{\Upsilon}
\newcommand{\intd}{Y_{Q}}
\newcommand{\Phiq}{\Phi_Q}
\newcommand{\BA}{\mbox{ABA}}

\begin{document}

\begin{center}
\vspace{24pt}
{ \LARGE Five loop Konishi from AdS/CFT}

\vspace{30pt}

{\large
Zolt\'an Bajnok$^{a,}$\footnote{e-mail: {\tt
bajnok@elte.hu}},\ \ 
\'Arp\'ad Heged\H us$^{b,}$\footnote{e-mail: {\tt hegedusa@rmki.kfki.hu}},\ \ 
Romuald A. Janik$^{c,}$\footnote{e-mail: {\tt ufrjanik@if.uj.edu.pl}}\\ and
Tomasz {\L}ukowski$^{c,}$\footnote{e-mail: {\tt tomaszlukowski@gmail.com}} 
}

\vspace{24pt}
{
\footnotesize
${}^a$ Theoretical Physics Research Group,
Hungarian Academy of Sciences,\\
1117 Budapest, P\'azm\'any s. 1/A, Hungary\\
${}^b$Research Institute for Particle and Nuclear Physics,
Hungarian Academy of Sciences,\\
H-1525 Budapest 114, POB 49, Hungary\\
${}^c$ Institute of Physics,
Jagiellonian University,\\
ul. Reymonta 4, 30-059 Krak\'ow, Poland\\
}

\vspace{48pt}

\end{center}

\begin{center}
{\bf Abstract}
\end{center}

\vspace{8pt}
\noindent
We derive the perturbative five loop anomalous dimension of the
Konishi operator in N = 4 SYM theory from the integrable string
sigma model by evaluating finite size effects using L{\"u}scher formulas
adapted to multimagnon states at weak coupling. In addition, we derive the five loop wrapping contribution for the $L=2$ single impurity state in the 
$\beta$ deformed theory, which may be within reach of a direct perturbative computation.
The Konishi expression exhibits two
new features - a modification of Asymptotic Bethe Ansatz quantization and sensitiveness to
an infinite set of coefficients of the BES/BHL dressing phase.
The result
satisfies nontrivial self-consistency conditions - simple transcendentality
structure and cancellation of $\mu$-term poles. It may be a testing ground for
the proposed AdS/CFT TBA systems.

\newpage

\section{Introduction}

The AdS/CFT correspondence \cite{adscft} states the equivalence of type IIB
superstrings on $AdS_5 \times S^5$ with the maximally supersymmetric
$\nn=4$ gauge theory in four dimensions. 
Local operators in gauge theory correspond to states in superstring theory.
In the planar limit, $N_c \to \infty$, multitrace operators factorize
essentially into independent single trace factors which correspond, on the
string side, to noninteracting strings. 
In this limit, one identifies the
spectrum of a type IIB superstring in $AdS_{5}\times S^{5}$ with the anomalous
dimensions of single trace operators in the maximally supersymmetric $\nn=4$
four-dimensional $SU(N_c)$ gauge theory.

This identification should hold for any
value of the 't Hooft coupling constant $\lm=g^2_{YM} N_c$. In the strong
coupling limit, and for operators with sufficiently large charges, the
corresponding strings may become classical/semi-classical and then the problem
becomes tractable by conventional methods. In contrast, once the coupling is
no longer large and/or we consider generic short operators the answer requires
us to consider the worldsheet QFT of the superstring on the quantum level. In
particular, if we would like to make contact with conventional perturbative
gauge theory calculations and explicitly test the correspondence or follow some
specific operator all the way from weak to strong coupling we would have to
quantize exactly the worldsheet QFT of the superstring which is a highly
nonlinear theory. 

Fortunately, the worldsheet QFT of the superstring in the $AdS_5 \times S^5$
background is integrable, which was first shown on the classical level in
\cite{Bena:2003wd}, and assuming that integrability also holds on the quantum
level (for which there are now many indications), one may use the theory of two-dimensional integrable quantum field theories to eventually quantize the theory in a quite explicit way 
for any value of $\lm$. The AdS/CFT correspondence thus allows us to use tools and
methods which exist only for \emph{two-dimensional} quantum field theories to
study the \emph{four-dimensional} $\nn=4$ gauge theory.

The theory of two-dimensional integrable quantum field theories is
well-developed
by now. The general strategy to determine their finite volume spectrum
goes as follows: First one considers the model in infinite volume.
The Hilbert space of asymptotic states is built up from noninteracting multiparticle
states which transform covariantly under the global symmetry algebra.
Time evolution is formulated in terms of the scattering matrix that
connects the initial and final multiparticle states \cite{IZ}. Once integrability
is assumed, there is no particle creation, and moreover any multiparticle scattering process is shown to factorize
into pairwise two particle scatterings. So the whole scattering information is contained in the $2\to2$ particle S-matrix.
The requirements of unitarity, crossing symmetry, invariance under the global
symmetry and the Yang-Baxter equation usually determines the scattering
matrix uniquely up to CDD type ambiguities \cite{ZamZam}. 
These are fixed through an analysis of the singularities of the scattering
matrix all of which must have a physical origin. Thus the appearing poles have to
correspond either to bound-states or to Coleman-Thun diagrams (anomalous
thresholds).  In this bootstrap solution we consider the bound-states
and the original particles on equal footing and determine the scattering
matrix of the bound-states from the scatterings on their individual
constituents \cite{Patrick}. Then the singularity structure of the bound-state scattering
matrix is analyzed and new bound-states are searched for. The so-called
bootstrap program is completed if all singularities of all the scattering
matrices are explained and then the theory is completely solved in
infinite volume. 

In the context of applications to the AdS/CFT correspondence, the infinite volume solution is not enough since one wants to describe closed strings and therefore consider the integrable quantum field theory on a cylinder of a given circumference related to a $U(1)_R$ charge of the given state. In contrast to conventional relativistic theories, \emph{all} (integer \footnote{See also \cite{Frolov:2009in} where it was shown that the meromorphicity of Y-system
leads to the quantization of the temperature of the mirror model.}) sizes of the cylinder are in fact relevant for the complete spectrum of the theory.

The finite volume
solution of the model can be achieved by systematically taking into
account the finite size effects due to the scatterings of particles.
The leading finite size effect of a multiparticle state comes from
the quantization of momenta. It is described by Bethe-Yang
equations and takes into account the scattering matrix in determining the
allowed momenta \cite{Luscher1}. It incorporates all polynomial corrections in the
inverse of the volume. In addition, there are exponentially small (L\"uscher)
corrections as well and their leading contributions come from the
polarization of the sea of virtual particles \cite{Luscher2}. For small volumes, these effects become dominant and one needs to perform a resummation of the virtual corrections, which sometimes can be carried out in the form of nonlinear integral equations.

Indeed, the exact description
of the finite volume ground state energy can be obtained from the
Thermodynamic Bethe Ansatz (TBA) \cite{ZAMTBA}. This is based on the fact that 
the contribution of the ground state
energy dominates the Euclidean partition function for large Euclidean
time, and that the same partition function can
be calculated by exchanging the role of space and time. One is left
with the determination of the partition function at finite temperature, but in the large volume limit where finite size effects are under
control. This method provides typically coupled integral equations
for pseudo energies which determine the ground state energy exactly.
In some circumstances a careful analytical continuation (in the volume
say) can provide integral equations for excited states as well \cite{PatRob1}. The
procedure is a numerical one and the resulting equations are only
conjectural which have to be further tested (but in all known cases they have passed all checks).

The analogous bootstrap solution of the $AdS_5 \times S^5$ worldsheet QFT of the light-cone quantized Green-Schwartz superstring is currently almost completed. Historically, the developments which led to it concentrated on the Bethe Ansatz part mostly on the gauge theory side \cite{Minahan:2002ve,Beisert:2003tq,Beisert:2003yb,BDS,K2} as well as for classical string solutions in $AdS_5 \times S^5$ \cite{KMMZ,AFS}. Later this was reformulated in the (spin-chain) S-matrix language in \cite{S,BS}, culminating in a proposal for the all loop Asymptotic Bethe Ansatz \cite{BS} the derivation of the (spin-chain) S-matrix from global symmetry properties \cite{B}. In fact, this derivation could be transformed in a verbatim way into a starting point for the bootstrap solution of the worldsheet QFT. This was necessary since the worldsheet QFT perspective was crucial in order to tackle a new `topological' class of Feynman graphs -- `wrapping diagrams' -- which went beyond the Bethe Ansatz in the spin chain guise\footnote{Note however the attempt to use the Hubbard model formalism for this purpose \cite{RSS}.}.

Let us therefore recall the basic steps of the 
bootstrap solution of the light-cone quantized Green-Schwartz string in $AdS_5\times S^5$. It
is classically integrable for any value of the light-cone momentum,
$P_{+}$ identified with the $U(1)_R$ charge $J$, which serves as the volume of the two-dimensional theory. A notable new feature of this theory is that it is \emph{not} relativistic invariant.
In the decompactification limit $J\to\infty$ the massive excitation
transform under the global symmetry algebra: the centrally extended
$su_c(2\vert2)^{2}$. This symmetry \cite{B}, together with unitarity and crossing
symmetry \cite{CROSS} completely fixes the scattering matrix\footnote{Taking into account some important subtleties \cite{ZF}.} including the dressing factor 
\cite{BHL,BES} modulo
CDD ambiguities \cite{Volin:2009uv}. The analysis of the pole structure revealed
an infinite tower of bound-states \cite{Dorey:2007an} whose scattering matrices
have been calculated as well \cite{Arutyunov:2008zt}. Double poles corresponding to anomalous thresholds were identified in 
\cite{DHM} and the related Coleman-Thun diagrams were found. As the physical region of the AdS S-matrix is not known the 
complete analysis of its singularity structure is rigorously not completed yet.  

In order to find the spectrum one has to pass to finite volume and consider the theory on a \emph{cylinder}. 
The corrections to the Asymptotic Bethe Ansatz can be identified with wrapping corrections \cite{AJK}, 
and the leading L{\"u}scher corrections for single \cite{JL} and multiparticle states  \cite{Bajnok:2008qj} 
have been derived for this (nonrelativistic) theory. In this case the space-time interchange necessary 
for formulating TBA leads to quite a different theory \cite{AFbound} with a distinct set of bound-states. 
By now TBA equations are also
developed by various groups
\cite{Bombardelli:2009ns,Gromov:2009bc,Arutyunov:2009ur}, under various analyticity assumptions, in particular about 
the analytically continued dressing phase. There are
conjectures also for excited states TBA equations \cite{Gromov:2009bc} but they have not
been tested yet beyond the leading perturbative order of wrapping corrections.

The formalism of the leading multiparticle L{\"u}scher corrections have been quantitatively tested in the case of the 4-loop anomalous dimension of the Konishi operator where the leading order wrapping
diagrams have been calculated perturbatively 
\cite{FSSZ,Velizhanin08}. The corresponding leading order L\"uscher calculation found an excellent agreement \cite{Bajnok:2008bm}. 
Subsequently wrapping interactions computed from L\"uscher corrections were found to be crucial 
for the agreement of some structural properties of twist two operators 
\cite{Bajnok:2008bm} with LO and NLO BFKL expectations \cite{Kotikov:2007cy}.

The aim of the present paper is to elaborate the L\"uscher correction
to next to leading order and by this to calculate the anomalous dimension
of the Konishi operator at five loops. Besides this explicit knowledge,
which can serve as a testing ground for excited state TBA equations,
we further test several issues of the formalism as well as a sizeable part of the BES/BHL dressing phase since the
calculation requires the knowledge of the analytically continued dressing
phase in the L\"uscher kinematics. It relies on the conjectured finite
size energy correction of multiparticle states, which at subleading
order contains corrections originating from the modification of ABA.
Although there is no perturbative gauge theory computation so far, the
internal consistency of our calculation provides enough confirmation
to believe in its correctness. In addition we consider the five loop subleading wrapping correction for a single impurity operator in the $\beta$ deformed theory, which may be within reach of a direct perturbative verification.

The paper is organized as follows: In section 2 we summarize the main
features of the 5-loop Konishi computation emphasizing the new phenomena which appear w.r.t. the previous 4-loop case, and the possible internal consistency checks of the computation. 
In section 3 we rederive
in a simple example the formalism of multiparticle L\"uscher correction.
We focus on a theory where the S-matrix does not depend on the difference
of the rapidities and derive TBA equation for the ground state. Excited
state energy levels are obtained by analytical continuation and,  by
analyzing their large volume asymptotics,  multiparticle L\"uscher
corrections are extracted. As they contain the analytically continued
scattering matrix, in Section 4 we determine the analytical continuation
of the dressing phase by two different methods. Section 5 is devoted
to the main computation of the anomalous dimension of the Konishi
operator. It starts with the calculation of the ABA. Then we turn
to the computation of the L\"uscher correction. It has two sources:
one comes from the modification of the ABA, the other corresponds to
the virtual particles circulating around the cylinder. Both terms
include an integration over the momentum of the mirror particles and
a summation over their spectrum. Integration is carried out by residues,
where, in comparison to the 4-loop case, we have to take into account an 
infinite tower of poles coming from the polygamma function part of the integrand.
The resulting expression is composed of rational and polygamma functions
which have to be summed over the bound-states. The technique developed
for the summation is explained in Section 6. Finally we give our conclusions in
Section 7. The paper is followed by two Appendices.
In the first we calculate the 5-loop anomalous dimension of a single
impurity operator, which acquires nontrivial wrapping corrections
in the $\beta$ deformed theory. In the second Appendix we explain
how to sum up terms containing polygamma functions.

\section{Main features of the 5-loop Konishi computation}

The Konishi operator $\tr \Phi_i^2$ is the simplest operator not protected by supersymmetry
which, thanks to its short size has proved to be a testing ground for the 
AdS/CFT correspondence. 
In most computations it
is more convenient to use a different representative of the same supermultiplet
which lies in the \slii sector
\eq
\tr (DZDZ) - \tr (ZD^2Z)
\eqx
Its anomalous dimension following from ABA is given by\footnote{See
section \ref{s.bakon} for a quick derivation.}
\eqn
\label{e.bethekon}
E_{ABA}&=&4+12g^2-48g^4+336 g^6-(2820+288\zeta(3))g^8 \\ \nonumber
&&\hspace{3cm}+(26508+4320\zeta(3)+2880\zeta(5))g^{10}+\ldots
\eqnx
However already at four loops there appears a contribution of wrapping graphs.
The four loop wrapping contribution was computed directly in perturbative gauge
theory \cite{FSSZ} using supergraph techniques and reconfirmed together with the
nonwrapping part using component Feynman graphs in \cite{Velizhanin08}. In \cite{Bajnok:2008bm}
the same result was computed from 
the string sigma model in $AdS_5 \times S^5$ and came from a L{\"u}scher type
F-term graph (see fig. 1) where a `virtual' particle was circulating in a loop. 

\begin{figure}
\centerline{\includegraphics[width=5cm]{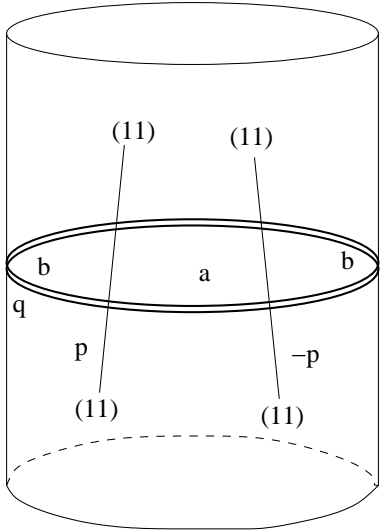}}
\caption{F-term of the L{\"u}scher correction corresponds to virtual particles
propagating around the worldsheet and scattering with the Konishi two-particle
state, $a$ and $b$ denote internal states of the virtual particle while $q$ is its `mirror' momentum.}
\end{figure}

In the L{\"u}scher corrections the virtual particle is strictly speaking
on-shell, however its kinematics are from the space-time interchanged theory
(so-called mirror theory). One can estimate its contribution to the magnitude of
the expression for the wrapping correction which is \cite{AJK}
\eq
e^{-2J \arcsinh \f{\sqrt{1+q^2}}{4g}} \longrightarrow \f{4^J g^{2J}}{(1+q^2)^J}
\eqx 
For the Konishi $J=2$, so this gives a factor of $g^4$. Together with the
so-called string frame phase factors \cite{ZF} which promote the string length
(light cone size of the worldsheet cylinder) to the `spin chain length' these
give another factor of $g^4$. Together the two factors give indeed $g^8$, making
this contribution to appear at 4-loop. We expect therefore that the correction
coming from two virtual particles should appear at least at order $g^{12}$ (or
if the string frame phase factors will also appear systematically this might be
even as late as $g^{16}$). Therefore it is expected that the 5-loop wrapping
part of the Konishi anomalous dimension will also appear from the leading
L{\"u}scher correction.
What makes this computation interesting, and what is the main motivation for our
study, is the fact that two new features which were absent in the 4-loop case
make their appearance here.

Firstly, the dressing factor of the S-matrix between the mirror particle and the
physical particles being the constituents of the Konishi state behaves like
$\exp(i g^2 \,phase)$, where the $phase$ involves contribution from an
\emph{infinite} set of BES coefficients. This is in stark contrast to the
behaviour of the dressing phase between physical particles where it behaves like
$\exp(i g^6 \,phase')$, and at higher orders the higher BES coefficients enter
only one by one. We will explicitly compute the dressing phase between mirror
and physical particles in section \ref{dressing.phase}.

Secondly, the virtual particle will also modify ABA quantization
condition. The reason that this effect appears only at 5-loop order is that the
momenta of the constituent particles get shifted from the ABA value
$p_{ABA}$ by a term of order $g^8$:
\eq
p=p_{ABA}+g^8 \dl_{w}p_{ABA}
\eqx
and hence their contribution to the energy coming from the dispersion relation
would appear only starting from order $g^{10}$
\eq
E(p)=\sqrt{1+16g^2 \sin^2 \f{p}{2}}=\sqrt{1+16g^2 \sin^2 \f{p_{ABA}}{2}} +4\sin p_{ABA}
\dl_{w}p_{ABA} g^{10}+\ldots
\eqx

The appearance of these two new effects is the main motivation for our
calculation. 

Of course the utility of this computation as a testing ground of the above two
phenomena depends on the ability of having a cross-check. For the Konishi
operator, a direct perturbative 5-loop computation seems to be beyond reach,
however there are some very stringent internal crosschecks of the eventual final
formula. 

Firstly, from the structure of perturbative gauge theory integrals we expect the
final answer to have a rather simple transcendentality structure -- a linear
combination of zeta functions (and possibly their products). Yet generically the
subexpressions which appear when performing the computation from the string
theory side are much more complicated like polygamma functions evaluated at
irrational complex arguments. A nontrivial cross check will be the cancellation
of these terms between the various parts of the computation, in particular
between the contribution of the dressing phase, the modification of Asymptotic Bethe Ansatz
and the direct F-term integral.

Secondly, purely from the integrable quantum field theory point of view we do not
expect to have a contribution of so-called $\mu$-terms due to the fact that at
weak coupling bound-states are much heavier than fundamental particles.
Typically $\mu$-terms arise from certain `dynamical' poles of the F-term
integrand.
A consequence of the vanishing of $\mu$-term contribution would be a
cancellation of the residues of those terms when summed over all bound-states.
This again necessitates a subtle cancellation between the residues coming from
the dressing phase, ABA modification and the F-term integrand. Therefore we may
use these two consistency checks as a test both of our formalism and of a huge
part of the BES dressing phase. Another motivation is the ongoing search and
proposals for the nonlinear integral equations/functional equations which would
exactly describe the spectrum for any size of the cylinder. The current
computation could then be used as a test of these proposals especially as the
modification of ABA quantization appears in a nontrivial way and
the dressing phase expression in this regime is quite complicated.

We will also consider a single impurity state with momentum $p=\pi$ which can be
considered as an analytical continuation of the twist-two operators considered
at 4-loop in \cite{Bajnok:2008qj}. It should also coincide with a physical state in the
$\beta$-deformed theory at $\beta=1/2$. Such states have been considered
perturbatively in \cite{Fiamberti:2008sn} and it may be possible to have a direct 5-loop
perturbative computation in this case. Here the modification of ABA
quantization is absent, however there is a contribution from the infinite set of
coefficients of the BES dressing phase. Therefore a direct perturbative
computation would be very interesting even for the single impurity case.

\section{Multiparticle L{\"u}scher formulas and ABA modification}

In this section we will review the formalism of multiparticle L{\"u}scher
corrections introduced in \cite{Bajnok:2008bm}. In order to avoid ambiguities
associated with the fact that the S-matrix for the $AdS_5 \times S^5$
superstring is a nontrivial function of both momenta and does not depend just on
the difference of rapidities as is the case for relativistic integrable quantum
field theories, we will consider the construction of the Thermodynamic Bethe
Ansatz for a theory with diagonal scattering but with an S-matrix \emph{without}
the difference property. Then we will consider the construction of excited state
TBA by analytical continuation along the lines of \cite{PatRob1}. We will
then extract the appropriate formulas for multiparticle L{\"u}scher corrections
by a large volume expansion recovering the expressions proposed in
\cite{Bajnok:2008bm}. However we hope that the present derivation will make their
origin clearer. We would like also to keep throughout the computation the
conventions for the S-matrix predominantly used for the $AdS_5 \times S^5$ case
which are different from the ones used usually for relativistic integrable
QFT's. 

\subsection{TBA for diagonal scattering}

The starting point for the derivation of the Thermodynamic Bethe Ansatz is the
consideration of the mirror theory with space and time interchanged. We denote
the momenta of the mirror theory by $\tilde p$ to make a clear distinction
compared to the momenta of the original sigma 
model what we denote by $p $. In comparing with the results of \cite{AJK} we
note that $\tilde p $ was denoted there by $p_{tba} $. 
We consider a theory with one species of particles scattering with the S-matrix
$S(\tilde p_{1},\tilde p_{2})$. 

In this theory the ABA takes the form\footnote{We use conventions used
in AdS/CFT literature.}
\eq
\label{e.bamirr}
e^{i\tilde p_{j}R}=\prod_{k:k\neq j} S(\tilde p_{j},\tilde p_{k})
\eqx
As is standard in TBA we are interested in the thermodynamic limit $R\to \infty$
of the free energy, where the physical size $L$ is identified with the inverse
temperature.
We will parametrize the momenta by $z$ (which may be identified with the
coordinate on the rapidity torus).
In the thermodynamic limit, the roots become very dense and can be substituted
by continuous densities of roots $\rho(z)$ and densities of holes $\rho_h(z)$
(unoccupied roots cf. \cite{ZAMTBA}). 

Taking the logarithm of (\ref{e.bamirr}), and the derivative w.r.t. $z$ we
obtain a relation between $\rho(z)$ and $\rho_h(z)$:
\eq
2\pi (\rho(z)+\rho_h(z)) = R \tilde p'(z) - \f{1}{i} \partial_z \log S(\tilde
p(z),\cdot) * \rho
\equiv R \tilde p'(z) -\phi*\rho
\eqx
In order to get a second equation necessary for solving for both densities,
we have to minimize the free energy which is given by
\eq
-RLf \equiv -LH+S  = -L\int \,  \tilde E(z) \rho(z) dz +  S[\rho,\rho_h]
\eqx
where $\tilde E $ is the mirror energy, while the entropy is 
\eq
S[\rho,\rho_h] = \int dz \left\{(\rho+\rho_h) \log(\rho+\rho_h) -\rho \log \rho
-\rho_h \log \rho_h \right\}
\eqx
It is convenient to introduce the pseudoenergy $\eps(z)$ which is related to the
densities of roots through
\eq
\f{\rho}{\rho+\rho_h} = \f{e^{-\eps}}{1+e^{-\eps}}
\eqx
and exremizing the free energy after taking into account the relation
\eq
\dl \rho+\dl \rho_h =-\f{1}{2\pi} \phi * \dl \rho
\eqx
coming from ABA, we obtain the final form of the TBA equation
\eq
\label{e.tba}
\eps(z)= L\tilde E(z) + \int \f{dw}{2\pi}\,
\phi(w,z)\log\left(1+e^{-\eps(w)}\right)
\eqx
We can now evaluate the thermal free energy of the mirror theory which gives the
\emph{physical} ground state energy of the original theory at zero temperature
but at finite size.
\eq
\label{e.entba}
E=L f= - \int \f{dz}{2\pi } \, \tilde p'(z) \log \left(1+e^{-\eps(z)}\right)
\eqx

\subsection{TBA for excited states by analytical continuation}

Initially, the Thermodynamic Bethe Ansatz could give information only on the
exact ground state energy, and it was not known how to extract similar
information for excited states in finite volume.
In \cite{PatRob1}, a version of the Thermodynamic Bethe Ansatz for excited
states was constructed by making an analytical continuation of the ground state
equations. The idea was that in deforming contours of integration in the TBA
equations, zeroes of $1+e^{-\eps(z)}$ would generate poles in the integrand
which could then be rewritten as additional source terms of the equations. Here
we will perform an analogous heuristic procedure, fixing the sign of the pole
contribution (which depends on the orientation of the contour w.r.t. the
singularities) in order to reproduce ABA results at large volume. Once
this is fixed, we will be able to unambiguously define leading corrections. Note
that we assume here that there are no $\mu$-terms so the calculation is
analogous to the Sinh-Gordon model in the relativistic case and so we will
assume that each physical particle is represented just by a single pole. 

We will need the relation between the mirror energies and momenta $(\tilde
E,\tilde p)$ and the physical ones $(E,p)$.
 These are defined by
\eq
\tilde E =ip \qqqq  \tilde p=iE
\eqx
which is fixed by taking the mirror integral to be for $z=x-\omega _2 /2$ with
$x$ real, and consequently identifying $z^-/z^+$ with $e^{-\tilde E}$. An opposite
choice is also possible.

Let us first consider the energy formula (\ref{e.entba}) integrated by parts:
\eq
E=\int \f{dz}{2\pi} \, \tilde p(z) \partial_z\log \left(1+e^{-\eps(z)}\right)
\eqx
Suppose that $1+e^{-\eps(z_{1,2})}=0$. Then the orientation has to be such that
the $\partial_z\log \left(1+e^{-\eps(z)}\right)$ contributes $-1$ by residues.
We thus obtain
\eq
\label{e.entwo}
E=E(z_1)+E(z_2)- \int \f{dz}{2\pi} \, \tilde p'(z) \log
\left(1+e^{-\eps(z)}\right)
\eqx
By the same mechanism an analogous contribution will arise from the TBA equation
(\ref{e.tba}), with the sign already fixed by the above considerations. We get
\eq
\label{e.tbatwo}
\eps(z)=L\tilde E +\log S(z_1,z)+\log S(z_2,z)+ \int \f{dw}{2\pi i}\,
(\partial_w \log S(w,z)) \log\left(1+e^{-\eps(w)}\right) 
\eqx
The multiparticle L{\"u}scher formulas will follow, for this theory, by
performing large volume (large $L$) expansion.

\subsection{Multiparticle L{\"u}scher formulas}

Firstly in order to compute the energy keeping the first exponential
corrections, we may neglect the integral term of (\ref{e.tbatwo}) when inserting
$\eps(z)$ into (\ref{e.entwo}). We thus get
\eqn
E &=& E(z_1)+E(z_2)- \int \f{dz}{2\pi} \, \tilde p' e^{-L \tilde E}
\f{1}{S(z_1,z) S(z_2,z)} \nonumber\\
&=& E(z_1)+E(z_2)- \int \f{d\tilde p}{2\pi}  \, e^{-L \tilde E} S(z,z_1)
S(z,z_2)
\eqnx
We recognize at once the F-term integral (with $q \equiv \tilde p$).

In order to complete the formula we need to self consistently fix the positions
of the poles $z_1$ and $z_2$. We will do it in two steps. First we neglect the
integral term of (\ref{e.tbatwo}) and impose for the rest $\eps(z_i)=i\pi+(2\pi
n)i$. We get
\eq
i \pi = \eps(z_1) =iLp_1 +i\pi + \log S(z_2,z_1)
\eqx
where we supposed that $S(z,z)=-1 $. This gives at once ABA
equation with our conventions
\eq
e^{i L p_1}=S(p_1,p_2)
\eqx

However it turns out that we have to be more precise in the determination of the
position of the roots and we have to include the integral term. In order to
define the quantization conditions we thus have to use
\eqn
\eps(z)&=&iLp(z)+ \log S(z_1,z)+\log S(z_2,z)+ \nonumber\\
&& + \int \f{dw}{2\pi i}\, \f{\partial_w S(w,z)}{S(w,z)} e^{-L \tilde E(w)} S(w,z_1)
S(w,z_2)
\eqnx 
The quantization conditions $\eps(z_i)=i\pi$ takes the form
\eqn
0&=&\underbrace{\log\{ e^{iL p_1} S(z_2,z_1)\}}_{BY_1} + \underbrace{ \int
\f{dw}{2\pi i }\,( \partial_w S(w,z_1)) S(w,z_2) e^{-L \tilde E(w)}}_{\Phi_1}
\\ 
0&=&\underbrace{\log\{ e^{iL p_2} S(z_1,z_2)\}}_{BY_2} + \underbrace{ \int
\f{dw}{2\pi i}\, S(w,z_1) (\partial_w S(w,z_2)) e^{-L \tilde E(w)}}_{\Phi_2}
\eqnx
Since the integrals are exponentially small we may solve these equations in
terms of corrections to  ABA giving
\eqn\label{BAmod.delatap1}
\f{\partial BY_1}{\partial p_1} \dl p_1 + \f{\partial BY_1}{\partial p_2} \dl
p_2 +\Phi_1 &=& 0 \\ \label{BAmod.delatap2}
\f{\partial BY_2}{\partial p_1} \dl p_1 + \f{\partial BY_2}{\partial p_2} \dl
p_2 +\Phi_2 &=& 0 
\eqnx
The final formula for the energy thus takes the form
\eq
\label{e.fin}
E=E(p_1)+E(p_2) + E'(p_1)\dl p_1+ E'(p_2)\dl p_2 - \int \f{dq}{2\pi}\, e^{-L
\tilde E} S(z,z_1) S(z,z_2)
\eqx
For the case of the $AdS_5\times S^5$ string sigma model we just have to replace
the product of the S-matrices by an appropriate supertrace. Thus the integrand
becomes essentially the transfer matrix. However the part coming from the
modification of ABA quantization does not have \emph{a-priori} such
a direct relation to the transfer matrix. Hence in the following we will
evaluate it directly from its definition using the S-matrices and their
derivatives.

\section{The dressing phase in the L\"uscher kinematics}\label{dressing.phase}

As we explained in section 2 one of the main sources of corrections which
appears at five loops is the contribution of the dressing phase. In
this section we present two calculations for its leading order part
in the L\"uscher kinematics.
By this we mean a kinematics which is relevant in calculating the
finite size correction, that is when the first argument is in the
mirror region $\vert x_{1}\vert<1$ while the second is in the physical
one $\vert x_{2}\vert>1$.

The dressing phase when both arguments are in the physical region,
($\vert x_{1,2}^{\pm}\vert>1$) can be written in the following form
\begin{equation}
\theta(x_{1},x_{2})=\chi(x_{1}^{+},x_{2}^{+})-\chi(x_{1}^{+},x_{2}^{-})-\chi(x_{
1}^{-},x_{2}^{+})+\chi(x_{1}^{-},x_{2}^{-})\label{eq:dressingphase}
\end{equation}
 where \eq
\chi(x_{1},x_{2})=-\sum_{r=2}^{\infty}\sum_{s>r}\frac{c_{r,s}(g)}{(r-1)(s-1)}
\left[\frac{1}{x_{1}^{r-1}x_{2}^{s-1}}-\frac{1}{x_{1}^{s-1}x_{2}^{r-1}}\right]
\eqx
The coefficients $c_{r,s}(g)$ have a convergent weak coupling expansion \cite{BES}
\begin{equation}
c_{r,s}(g)=2\cos\left(\frac{\pi}{2}(s-r-1)\right)(r-1)(s-1)\int_{0}^{\infty}
dt\frac{J_{r-1}(2gt)J_{s-1}(2gt)}{t(e^{t}-1)}\label{eq:BES}\end{equation}
As was shown in \cite{DHM} the dressing phase has also an alternative double integral representation, 
which is 
suitable for analytical continuations: \begin{eqnarray}
\chi(x_{1},x_{2})=i\oint_{C_{1}}\frac{dw_{1}}{2\pi
i}\frac{1}{w_{1}-x_{1}}I_{2}(w_{1},x_{2})\label{eq:DHM}\end{eqnarray}
where
\begin{eqnarray}
 I_{2}(w_{1},x_{2})=\oint_{C_{1}}\frac{dw_{2}}{2\pi
i}\frac{1}{w_{2}-x_{2}}\log\frac{\Gamma(1+ig(w_{1}+w_{1}^{-1}-w_{2}-w_{2}^{-1}))
}{\Gamma(1-ig(w_{1}+w_{1}^{-1}-w_{2}-w_{2}^{-1}))} \end{eqnarray}
and the integrations go over the unit circles. In the next subsections
we focus on the weakly coupled regime, review the dressing phase for physical
particles in this regime and calculate the analytical
continuation for both representations in the L\"uscher kinematics.

\subsection{BES dressing phase for physical particles}

In the $g\to0$ limit we can expand the Bessel function as \eq
J_{n}(2gt)=\frac{g^{n}t^{n}}{n!}(1-\frac{t^{2}g^{2}}{n+1}+\dots)\eqx
 and perform the integration in (\ref{eq:BES}). This provides the
leading order behaviour of $c_{r,s}(g)$: 
 \begin{eqnarray}
c_{r,s}(g) & = &
\frac{2\cos(\frac{\pi}{2}(s-r-1))}{(r-2)!(s-2)!}g^{r+s-2}\bigl[
(r+s-3)!\zeta(r+s-2)+\label{eq:weakc}\\
 &  & \hspace{4cm}-\frac{g^{2}}{rs}(r+s)(r+s-1)!\zeta(r+s)+\dots\bigr]\nonumber
\end{eqnarray}
In calculating the L\"uscher correction we have to analyze the dressing
phase (\ref{eq:dressingphase}) for the case when the parameters $x_{1},x_{2}$
in the $g\to0$ limit behave as \eq
x_{1}^{+}=\frac{q+iQ}{2g}+O(g)\quad;\quad
x_{1}^{-}=\frac{2g}{(q-iQ)}+O(g^{3})\quad;\quad x_{2}^{\pm}=\frac{2u\pm
i}{2g}+O(g^{3})\eqx

Let us start with the case when both parameters are in the physical
kinematics $x_{1,2}=a_{1,2}g^{-1}+O(g)$. Then the leading order behaviour
comes from the term $c_{2,3}(g)$ and is given by 
\begin{equation}
\chi(x_{1},x_{2})=-(2g^{6}\zeta(3)+20g^{8}\zeta(5))\frac{a_{1}-a_{2}}{a_{1}^{2}
a_{2}^{2}}+O(g^{10})\label{eq:weakdressing}
\end{equation}
The next coefficients $c_{2,5}(g)$ and $c_{3,4}(g)$ only enter at $g^{10}$
order.
For physical particles, the dressing phase starts at order $g^{6}$
and this is what we have to use
in the calculation of the ABA. 

Let us now describe the regime where one of the particles is in the L{\"u}scher
('mirror') kinematic which is relevant for the
five loop wrapping computation.

\subsection{BES dressing phase in the L\"uscher kinematics}

Suppose now that the first argument is in the mirror kinematics
$x_{1}=a_{1}^{-1}g+O(g^{3})$
while we keep the second in the physical one $x_{2}=a_{2}g^{-1}+O(g)$.
Since $c_{r,s}(g)$ scales as $g^{r+s-2}$ while $x_{1}^{1-r}x_{2}^{1-s}$
as $g^{s-r}$ the leading order contribution comes from the $r=2$
part of the first sum: \eq
\chi(x_{1},x_{2})=\sum_{s>2}\frac{c_{2,s}(g)}{(s-1)}\frac{1}{x_{1}^{s-1}x_{2}}
+O(g^{4})\eqx
Clearly this is in stark contrast to the case of the ABA as in this
L\"uscher kinematics infinite number of coefficients contribute.
Using the leading order term of the explicit weak coupling expansion
of $c_{2,s}$ determined in (\ref{eq:weakc}) we obtain \begin{eqnarray*}
\chi(x_{1},x_{2}) & = &
\frac{g^{2}}{a_{2}}\sum_{s>2}2\cos(\frac{\pi}{2}(s-3))a_{1}^{s-1}\zeta(s)+O(g^{4
})\\
 & = & \frac{g^{2}}{a_{2}}(S_{1}(-ia_{1})+S_{1}(ia_{1}))+O(g^{4})\end{eqnarray*}
where $S_{1}(n)=\sum_{k=1}^{n}\frac{1}{k}$ is the harmonic number
which has an analytical continuation in terms of the digamma function
$\psi(x)=\frac{d\log\Gamma(x)}{dx}$ as \begin{eqnarray}
\chi(x_{1},x_{2}) & = &
\frac{g^{2}}{a_{2}}(2\gamma_{E}+\psi(1-ia_{1})+\psi(1+ia_{1}))+O(g^{4})\nonumber\\
 & = &\label{BESfinal}
\frac{g^{2}}{a_{2}}(2\gamma_{E}+\psi(-ia_{1})+\psi(ia_{1}))+O(g^{4})
\end{eqnarray}
where $\gamma_{E}$ is the Euler constant. Thus for sufficiently small
$a_{1}$ the original BES dressing phase provided a convergent expansion
for the leading order $g\to0$ behaviour in the L\"uscher kinematics.
The resulting expression then can be analytically continued for any
$a_{1}$. Let us check that this is the right analytical continuation
by calculating the weak coupling expansion of the analytically continued
dressing phase from the DHM integral representation.

\subsection{BES dressing phase in the L\"uscher kinematics from the DHM integral
formula}

The proper analytical continuation of the dressing phase (\ref{eq:DHM})
into the L\"uscher kinematics, compatible with crossing symmetry,
was determined in \cite{AF}: \eq
\chi(x_{1},x_{2})=i\oint_{C_{1}}\frac{dw_{1}}{2\pi
i}\frac{1}{w_{1}-x_{1}}I_{2}(w_{1},x_{2})-i I_{2}(x_{1},x_{2})\eqx
As in the previous subsection we are interested in the weak coupling
expansion of $\chi$ in the L\"uscher kinematics so we take
$x_{1}=a_{1}^{-1}g+O(g^{3})$
and $x_{2}=a_{2}g^{-1}+O(g)$. Observe that since $\vert x_{2}\vert>1$
the contour never encircles $x_{2}$. Let us focus on the second term.
We change the integration variable to $u_{2}=gw_{2}$ and write \eq
I_{2}(x_{1},x_{2})=\oint_{C_{g}}\frac{du_{2}}{2\pi
i}\frac{1}{u_{2}-a_{2}}\log\frac{\Gamma(1+ig(x_{1}+x_{1}^{-1})-iu_{2}-ig^{2}u_{2
}^{-1}))}{\Gamma(1-ig(x_{1}+x_{1}^{-1})+iu_{2}+ig^{2}u_{2}^{-1}))}\eqx
Since the integration for $u_{2}$ goes over a shrinking circle of
radius $g\to0$ we pick up the contributions of the poles at $u_{2}=0$
only (not at $a_{2}$). From the expansion of the $\Gamma$ function
in $g^{2}u_{2}^{-1}$ one can observe that higher order poles contribute
at higher orders in $g^{2}$. Thus when we focus on the leading order
behaviour we are allowed to keep the first order pole: \eq
I_{2}(x_{1},x_{2})=i\left(\frac{g^{2}}{a_{2}}\right)\left[\psi(1+ig(x_{1}+x_{1}^{
-1}))+\psi(1-ig(x_{1}+x_{1}^{-1}))\right]+O(g^{4})\eqx
Taking into account that $x_{1}=a_{1}^{-1}g+O(g^{3})$, the leading
order part turns out to be \eq
I_{2}(x_{1},x_{2})=i\left(\frac{g^{2}}{a_{2}}\right)\left[\psi(1+ia_{1}
)+\psi(1-ia_{1})\right]+O(g^{4})\eqx
In addition to this term we also have a term of the form \eq
I_{1}(x_{1},x_{2})=i\oint_{C_{1}}\frac{dw_{1}}{2\pi
i}\frac{1}{w_{1}-x_{1}}I_{2}(w_{1},x_{2})\eqx
Since $x_{1}=a_{1}^{-1}g+O(g^{3})$, it lies inside the integration
contour. We need a convergent small coupling expansion. For this we
change the integration variable to $u_{1}=w_{1}g^{-1}$ and close
the contour around $\infty$. The leading oder result comes only from
the pole at $\infty$ and reads as: \eq
I_{1}(x_{1},x_{2})=\left(\frac{g^{2}}{a_{2}}\right)2\gamma_{E}+O(g^{4})\eqx
 In order to determine the leading order behaviour of $\chi$ we have
to combine $I_{1}$ and $I_{2}$. The result provides the leading
order asymptotics of the dressing phase in the L\"uscher kinematics
\eq
\chi(x_{1},x_{2})=\left(\frac{g^{2}}{a_{2}}\right)\left[2\gamma_{E}+\psi(1+ia_{1
})+\psi(1-ia_{1})\right]+O(g^{4})\eqx
 which is consistent with what we obtained from the BES dressing phase.
This is the main result of this section. We will need to take into account
this expression when we calculate the five loop anomalous dimension of
the Konishi operator in the next section.

\section{The Konishi computation}

In this section, which is the main part of the paper we describe and
evaluate the various ingredients which together contribute to the 5-loop
wrapping correction. For completeness we first describe the evaluation of
the Asymptotic Bethe Ansatz (ABA) contribution to the anomalous dimension and
then proceed to compute the various parts of the wrapping correction,
summarizing the previous 4-loop result in the process. 

\subsection{Asymptotic Bethe Ansatz for the Konishi operator}
\label{s.bakon}

In this subsection we calculate the scaling dimension of the Konishi
operator from the ABA. In principle the complete scaling dimension has the
form \eq
\Delta(g)=\Delta_{ABA}(g)+\Delta_{w}(g)\eqx
where the first term contains the result of the ABA, while the second
the wrapping contributions. In doing the calculation we choose the
representative for the Konishi operator in the $sl_{2}$ sector. ABA
amounts to calculate in this case the energy of the two particle state
in volume $L=2$ with momentum $p$ and $-p$ which satisfy the equation
\eq
e^{ipL}=S(p,-p)=S^{sl(2)}(p,-p)e^{2i\theta(x(p),x(-p))}\eqx
where $S^{sl(2)}(p,-p)$ is the scattering matrix in the $sl_{2}$
sector which reads as 
\begin{equation}
S^{sl(2)}(x^\pm(p),x^\pm(-p))=\f{x^-(p)-x^+(-p)}{x^+(p)-x^-(-p)} 
\f{1-\f{1}{x^+(p) x^-(-p)}}{1-\f{1}{x^-(p) x^+(-p)}}
\end{equation}
 and $x^\pm(p)= \frac{1}{4g}(\cot(\frac{p}{2})\pm i)(1+E(p))$. 
We can solve this equation perturbatively. We take the Ansatz
$p=\sum_{i}p^{(i)}g^{2i}$
and systematically expand both the scalar and the dressing part
in increasing orders of $g$. The first two orders can be determined
without the dressing phase. These solutions then can be put into formula
(\ref{eq:weakdressing}) to determine the leading contributions of the 
dressing phase. Finally we find the solution for $p$ at the required
order \begin{eqnarray}
\label{e.p}
p&=&\frac{2\pi}{3}-\sqrt{3}g^{2}+\frac{9\sqrt{3}}{2}g^{4}-24\sqrt{3}(1+\zeta(3))g^
{6}\\
&&\hspace{5cm} +\frac{\sqrt{3}}{4}(671+960(\zeta(3)+\zeta(5)))g^{8}+O(g^{10})\nonumber \end{eqnarray}
 This momentum then determines the ABA part of the scaling dimension
as 
\begin{eqnarray*}
\Delta_{ABA}(g) & = & 2E(p)=4+12g^{2}-48g^{4}+336g^{6}-(2820+288\zeta(3))g^{8}\\
 &  &
\hspace{3cm}+(26508+4320\zeta(3)+2880\zeta(5))g^{10}+O(g^{12})
\end{eqnarray*}
The wrapping part of the anomalous dimension starts as 
\eq
\Delta_{w}(g)=\Delta_{w}^{(8)}g^{8}+\Delta_{w}^{(10)}g^{10}+O(g^{12})
\eqx
In \cite{Bajnok:2008bm} the 4-loop part has been evaluated to give
\eq
\Delta_{w}^{(8)} = 324 +864 \zeta(3) -1440 \zeta(5)
\eqx
and the aim of the rest of this paper is to calculate the wrapping part at
5-loop order $\Delta_{w}^{(10)}$. Our final result is
\begin{equation}
\Delta_w^{(10)} =-11340+2592\zeta(3) -5184\zeta(3)^2 -11520\zeta(5)
+30240\zeta(7)
 \end{equation}

\subsection{The general structure of the wrapping correction}

There are two sources of the L{\"u}scher correction for the Konishi operator
which have to be taken into account when we calculate the 5-loop anomalous
dimension (recall formula (\ref{e.fin})). 

\subsubsection*{Modification of Asymptotic Bethe Ansatz quantization}

First of all the leading contribution which comes from
the modification of ABA is of order $g^{10}$. This has the form
\eq
\label{e.enwba}
\Delta_w^{ABA}=E'(p_1)\dl p_1+E'(p_2)\dl p_2
\eqx
where $p_1=-p_2\equiv p$ is the solution (\ref{e.p}) of the ABA, while $\dl p_i$
are the shifts due to the virtual corrections. For the 5-loop result it will be
enough to take $p=2\pi/3$ in this formula.

The shifts $\dl p_i$  for the Konishi operator can be found from
relations (\ref{BAmod.delatap1}) and (\ref{BAmod.delatap2}) which in our case
take the form
\begin{eqnarray*}
\frac{5i}{2} \delta p_1-\frac{i}{2} \delta p_2+\Phi_1 &=& 0\\
-\frac{i}{2} \delta p_1+\frac{5i}{2} \delta p_2+\Phi_2 &=& 0 
\end{eqnarray*}
where $\Phi=\Phi_1=-\Phi_2$ and $\Phi$ for the Konishi operator is as follows
\begin{eqnarray}\label{Phi}
i \Phi&=&\sum_{Q}\int_{-\infty}^{\infty}\frac{dq}{2\pi}\left(\frac{
z^{-}}{z^{+}}\right)^{L}\sum_{b}(-1)^{F_{b}}\left[(\partial_q S_{Q-1}(q,u_{i}))S_
{Q-1}(q,u_{ii})\right]_{b(11)}^{b(11)} \\ \nonumber
 & \equiv&
\sum_{Q=1}^{\infty}\int_{-\infty}^{\infty}\!\frac{dq}{2\pi} \,\Phiq(q,u)
\end{eqnarray}
Solving the equations for $\delta p_1$ and $\delta p_2$ we found that 
\begin{equation*}
\delta p_1=-\delta p_2=\frac{i}{3}\Phi
\end{equation*}
This means that the momenta of the Konishi constituents are shifted in opposite
directions by the same factor. It leads to the vanishing of the total momentum after
ABA modification as it should be. 

When plugged into the dispersion relation it gives us the energy shift
(\ref{e.enwba}) due to
the Asymptotic Bethe Ansatz modification as
\begin{equation}
\Delta_w^{ABA}=\frac{4}{\sqrt{3}}\Phi
\end{equation}
We will evaluate $\Phi$ explicitly later in the section.

\subsubsection*{The F-term integral}

The second
contribution to the 5-loop result comes from the expansion of formula
\cite{Bajnok:2008bm}
\begin{equation}
\Delta^F_{w}=-\sum_{Q=1}^{\infty}\int_{-\infty}^{\infty}\!\frac{dq}{2\pi}
\left(\frac{
z^{-}}{z^{+}}\right)^{L}\sum_{b}(-1)^{F_{b}}\left[S_{Q-1}(q,u_{i})S_
{Q-1}(q,u_{ii})\right]_{b(11)}^{b(11)}\label{wrapping}
\end{equation}
up to subleading terms. We parametrize the virtual particle using the mirror
momentum $q$ as in \cite{Bajnok:2008bm} (see also (\ref{zplusminus})), 
while for the physical particles we use rapidities
as in \cite{Bajnok:2008qj}.
Hence $u_i=-u_{ii} \equiv u$ are the rapidities of the two particles forming the
Konishi state. They are related to the momentum $p$ as 
\eq
u(p)=\frac{1}{2}\cot(\frac{p}{2})\sqrt{1+16g^2 \sin^2(\frac{p}{2})}
\eqx
The rapidity variable for the Konishi state is therefore given as
\eq
\label{e.uexp}
u=u_{0}+u_{2}g^{2}+\ldots=\frac{1}{2\sqrt{3}}+\frac{4}{\sqrt{3}}g^{2}+\ldots
\eqx
To simplify our further considerations we rewrite (\ref{wrapping}) in the form 
\begin{eqnarray}
\label{e.defs}
\Delta^F_{w} & = &
-\sum_{Q=1}^{\infty}\int_{-\infty}^{\infty}\!\frac{dq}{2\pi}\Big(\frac{z^{-}}{z^
{+}}\Big)^{L}\sca(q,Q,u) \dress(q,Q,u) \mat(q,Q,u)^2  \\ \nonumber
 & = &
-\sum_{Q=1}^{\infty}\int_{-\infty}^{\infty}\!\frac{dq}{2\pi}\intd(q,u)
\end{eqnarray}
where we split the contribution coming from the S-matrix into three parts:
the scalar, the dressing and the matrix part. Each of those parts involves a
product over contributions of the two particles forming the Konishi state.

It is convenient to perform the weak coupling expansion of this formula in two
steps. First we keep $u$ as a variable of order 1 and expand all other sources of $g$
dependence. Then we take into account the fact that $u$ itself is $g$ dependent
and compute the $g$ dependent terms due to (\ref{e.uexp}).

All ingredients in formula (\ref{e.defs}) can be expanded into a series in
$g^{2}$, keeping $u$ fixed, as
\begin{eqnarray*}
\Big(\frac{z^{-}}{z^{+}}\Big)^{2} & = &
g^{4}\ex^{(4)}(q,Q)+g^{6}\ex^{(6)}(q,Q)+\ldots\\
\sca(q,Q,u) & = & \sca^{(0)}(q,Q,u)+g^{2}\sca^{(2)}(q,Q,u)+\ldots\\
\dress(q,Q,u) & = & 1+g^{2}\dress^{(2)}(q,Q,u)+\ldots\\
\mat(q,Q,u) & = &
g^{2}\mat^{(2)}(q,Q,u)^2+g^{4}\mat^{(4)}(q,Q,u)+\ldots
\end{eqnarray*}
Now we will factor out the leading piece
\eq
\label{e.leading}
\intd^{(8)}(q,u)g^{8} = 
\sca^{(0)}(q,Q,u)\mat^{(2)}(q,Q,u)^2\ex^{(4)}(q,Q,u)g^{8}
\eqx
and rewrite the first subleading term as a sum of contributions coming from the
matrix part, scalar part, the exponential term and the dressing factor:
\begin{eqnarray*}
\intd^{(10)}(q,u) & = &
\intd^{(8)}(q,u)\Big[2\frac{S_{\boxplus}^{(4)}(q,Q,u)}{S_{\boxplus}^{(2)}(q,Q,u)
}+\frac{S_{0}^{(2)}(q,Q,u)}{S_{0}^{(0)}(q,Q,u)}\\
 &  &\hspace{5cm}+
\frac{\ex^{(6)}(q,Q)}{\ex^{(4)}(q,Q)}+\dress^{(2)}(q,Q,u)\Big]
\end{eqnarray*}
In the following we will calculate all these contributions one by one.

Since $\intd^{(10)}(q,u)$ is already at order $g^{10}$, we may safely set here
$u=\frac{1}{2\sqrt{3}}$.
\eq
\intd^{(10)}(q,u)g^{10}= \intd^{(10,0)}(q) g^{10} + \oo{g^{12}}
\eqx
where $\intd^{(10,0)}(q)=\intd^{(10)}(q,u_0)$.

As mentioned earlier we also have to take into
account that the rapidity $u$ is $g$ dependent by itself. Clearly the
only part that will contribute at order $g^{10}$ is the expansion of
(\ref{e.leading})
\eq
\intd^{(8)}(q,u)g^{8} =
\intd^{(8,0)}(q) g^8 +\intd^{(8,2)}(q) g^{10}+\ldots
\eqx
with $\intd^{(8,0)}(q)=\intd^{(8)}(q,u_0)$.
To summarize, the wrapping correction at 4- and 5-loop respectively take the
form
\eqn
\Delta_{w}^{(8)} &=& -\sum_{Q=1}^\infty \int_{-\infty}^\infty \f{dq}{2\pi}
\intd^{(8,0)}(q) \\
\Delta_{w}^{(10)} &=& -\sum_{Q=1}^\infty \int_{-\infty}^\infty \f{dq}{2\pi}
\left( \f{4}{\sqrt{3}} \Phiq(q)+ \intd^{(10,0)}(q) + \intd^{(8,2)}(q) \right)
\eqnx

\subsubsection*{The leading contribution and 4-loop wrapping}

The leading expansion of integrand $\intd(q,u)$ is known to be proportional to
$g^{8}$ and was found in \cite{Bajnok:2008bm} as 
\begin{eqnarray*}
\intd^{(8)}(q,u)g^{8} & = &
\sca^{(0)}(q,Q,u)\mat^{(2)}(q,Q,u)^2\ex^{(4)}(q,Q,u)g^{8}\\
 &  &
\hspace{-4cm}=\frac{16384g^{8}Q^{2}(-1+q^{2}+Q^{2}-4u^{2})^{2}}{(q^{2}+Q^{2})^{4
}((q+i(Q+1))^{2}-4u^{2})((q+i(Q-1))^{2}-4u^{2})}\times\\
 & \times &
\frac{1}{((q-i(Q-1))^{2}-4u^{2})((q-i(Q+1))^{2}-4u^{2})}
\end{eqnarray*}

When expanded further using the fact that the rapidity variable $u$
is $g^{2}$-dependent (\ref{e.uexp})
 it can be rewritten as \eq
\intd^{(8)}(q,u)=\intd^{(8,0)}(q)+g^{2}\intd^{(8,2)}(q)\eqx
 where $\intd{}^{(8,0)}(q)$ is the only contribution which is relevant
for the 4-loop calculations. 
\eq
\Delta_w^{(8)} = -\sum_{Q=1}^\infty \int_{-\infty}^\infty \intd{}^{(8,0)}(q)
\eqx
The integrand is a rational function and hence the integral can be carried out
by residues. The
poles are of two kinds. Firstly there is a fourth order pole at $q=iQ$ which we
call `kinematical' pole since it comes just from the exponential terms. The
remaining poles come from the S-matrix and are `dynamical' i.e. 
$s$ and $t$ channel poles. They would be associated to possible $\mu$-terms.
Since, as argued in \cite{Bajnok:2008bm}, $\mu$-terms should not be present
at weak coupling, we expect that the residues of the dynamical poles should sum
up to zero after summation over all $Q$. We can indeed verify that this is the
case. Hence the 4-loop wrapping correction to the Konishi anomalous dimensions
can be obtained just by summing the residue of the kinematical pole only:
\eq
\Delta_w^{(8)}=-2\pi i \sum_{Q=1}^\infty \mathop{\res}_{q=iQ} \intd^{(8,0)}(q)  
=324+864\zeta(3)-1440\zeta(5)
\eqx
We expect that the same cancellation of `dynamical' poles will also hold at
5-loop. Indeed this will be a nontrivial internal consistency check of our
calculation.

\subsection{The 5-loop integrand}
The full integrand which will be relevant for 5-loop calculations consists of
the subleading contribution of $\intd^{(8)}(q,u)$ (namely $\intd^{(8,2)}(q)$),
the leading contribution of $\intd^{(10)}(q,u)$ (namely $\intd^{(10,0)}(q)$
which comes from the matrix, scalar, exponential part and from the dressing
phase) and
the leading contribution $\Phiq(q)$ coming from the ABA modification.
We will now evaluate one by one the individual contributions.

\subsubsection*{Matrix part}
The subleading matrix part of integrand $S_{\boxplus}^{(4)}(q,Q,u)$ for the
Konishi operator can be evaluated using the formulas (78)-(82) from
\cite{Bajnok:2008bm}. When normalized with respect to leading term
$S_{\boxplus}^{(2)}(q,Q,u)$ it stands as
	\begin{eqnarray*}
	&&\frac{S_{\boxplus}^{(4)}(q,Q,u)}{S_{\boxplus}^{(2)}(q,Q,u)}= \frac{16
q (1 - i q - Q - 4 u^2)}{(q + i Q) ((q - i (Q-1))^2 - 
    4 u^2) (1 + 4 u^2)} - 
	\frac{16}{(1 + 4 u^2)^2} \\
	&&\hspace{1cm}+\frac{16 ( q^2 + Q^2-1)}{(q^2 + Q^2) ( q^2 + Q^2-1 - 4 u^2))} +
\frac{4 (-5 + 5 q^2 + 4 Q + 5 Q^2 - 4 u^2)}{(q^2 + Q^2) (1 + 4 u^2)}
	\end{eqnarray*}

\subsubsection*{Scalar part}
The scalar part of the integrand can be evaluated using the S-matrices of the
scattering of the \slii bound-state constituents  $(z_1^-,z_1^+)$, \ldots,
$(z_Q^-,z_Q^+)$ with the fundamental magnon $(x^-,x^+)$:
\eq
\label{e.sl2fuse}
\sca(z^\pm,x^\pm)=\prod_{i=1}^Q S^{sl(2)}(z_i^\pm,x^\pm)
\eqx
with
\begin{equation*}
S^{sl(2)}(z^\pm,x^\pm)=\f{z^--x^+}{z^+-x^-} \f{1-\f{1}{z^+ x^-}}{1-\f{1}{z^-
x^+}}
\end{equation*}

In order to calculate (\ref{e.sl2fuse}) we use for $z_1^-$ the value $z^-$ and
for $z_Q^+$ the value $z^+$ where
\eq \label{zplusminus}
z^\pm = \f{Q}{4g} \left( -\sqrt{1+\f{16g^2}{Q^2+q^2}}\mp 1\right)
\left(-\f{q}{Q}-i\right)
\eqx
whereas $z_k^+$ we take from
\begin{equation*}
z_k^+=\f{1}{2} \left( z_k^-+\f{1}{z_k^-} +\f{i}{g}+ \sqrt{\left(
z_k^-+\f{1}{z_k^-} +\f{i}{g} \right)^2-4 } \right)
\end{equation*}
The intermediate $z_k^-$ are determined from the pole condition
$z_{k}^-=z_{k-1}^+$. In principle we could choose different signs
in front of the square roots. This amounts to choosing a different
representative for the constituents of the Q bound-state. 
It was shown however in \cite{AF} that all of the  different constituents
lead to the same S-matrix after analytical continuation. 
Our choice is technically the simplest one and 
leads to the expansion of the bound-state parameters up to the second order as
\begin{equation}
z_{k}^{+}=\frac{2ik+q-iQ}{2g}+\frac{2g}{q-iQ}-\frac{2g}{2ik+q-iQ}+\frac{2g}{q+iQ
}
\end{equation}

The result for the leading order scalar part of the integrand is then
\begin{eqnarray*}
\sca^{(0)}(q,Q,u)&=&\frac{((q - i( Q-1))^2 - 4 u^2) (1 + 4 u^2)^2}{((q -i (
Q+1))^2 - 
   4 u^2) ((q + i ( Q+1))^2 - 4 u^2) }\times\\
   &&\hspace{4cm}\times\frac{1}{((q + i ( Q-1))^2 - 4 u^2)}
\end{eqnarray*}
while the subleading 
one can be split into two parts: a rational part
\begin{eqnarray*}
&&\frac{S_{0rat}^{(2)}(q,Q,u)}{\sca^{(0)}(q,Q,u)}=-\frac{32}{(i q + Q) (1 +  4
u^2)} + \frac{32}{(1 + 4 u^2)^2}+\\
&&+\frac{8}{q^2 + Q^2} \Big( \frac{2 q (q - i ( Q-1))}{(q - i ( Q-1))^2 - 4 u^2}
- 
\frac{  2 q (q + i ( Q-1))}{(q + i ( Q-1))^2 - 4 u^2}+ \\
&&- \frac{ 2 q (q - i ( Q+1))}{(q - i ( Q+1))^2 - 4 u^2} - 
\frac{    2 q (q + i ( Q+1))}{(q + i ( Q+1))^2 - 4 u^2} - \frac{ 2 (q^2 + Q ^2 +
2 Q) }{1 + 4 u^2}\Big) 
\end{eqnarray*}
and one which contains polygamma functions
\begin{equation}
\frac{S_{0\psi}^{(2)}(q,Q,u)}{\sca^{(0)}(q,Q,u)}=\frac{16}{1 + 4 u^2} \Big(
\psi\big( \frac{1}{2}(-i q - Q)\big) - 
  \psi\big(\frac{1}{2} (-i q + Q)\big) \Big)
\end{equation}
\subsubsection*{Exponential part}
The leading and subleading term of the exponent which appears in $\intd$ is found
to be
\begin{equation*}
\Big(
\frac{z^-}{z^+}\Big)^2=\frac{16g^4}{(q^{2}+Q^{2})^{2}}\left[1-g^{2}\frac{16}{(q^
{2}+Q^{2})}\right]
\end{equation*}
where $z^+$ and $z^-$ is taken from (\ref{zplusminus}).
\subsubsection*{Dressing part}
The dressing part can be found from formula (\ref{BESfinal}) as
\begin{equation}
\dress^{(2)}(q,Q,u)=-\frac{32 }{1 + 4 u^2}\Big( \gamma_E + \frac{1}{2} \psi\big(
 \frac{1}{2}(-i q - Q)\big) + 
    \frac{1}{2} \psi\big( \frac{1}{2} (i q + Q)\big) \Big)
\end{equation}

\subsubsection*{ABA modification}

To make our notation more compact we rewrite the leading order Asymptotic Bethe Ansatz
modification formula into the form
\begin{equation}
\Phi^{(8)}=
\sum_{Q=1}^{\infty}\int_{-\infty}^{\infty}\!\frac{dq}{2\pi} \,\Phiq(q,u)
=\sum_{Q=1}^{\infty}\int_{-\infty}^{\infty}\!\frac{dq}{2\pi}\intd^{(8)}(q,
u)\BA(q,Q,u)
\end{equation}
In order to find $\BA(q,Q,u)$ it is enough to use formulas (78)-(82) from
\cite{Bajnok:2008bm}. Following (\ref{Phi}) we have to take the derivative of
S-matrix elements with respect to $q$ and then calculate the supertrace. The final
result we obtained is
\begin{eqnarray*}
&&\BA(q,Q,u)=-\frac{2 q}{q^2 + Q^2} + \frac{1}{-i - q - i Q + 2 u} + \frac{1}
 {i - q -i Q + 2 u} \\
&&\hspace{2cm}+ \frac{1}{-i - q + i Q + 2 u} + \frac{1}{
 i - q + i Q + 2 u} - \frac{2 (q + 2 u)}{1 - q^2 - Q^2 + 4 u^2}
\end{eqnarray*}

\subsubsection*{The contribution $\intd^{(8,2)}(q)$}

As mentioned in the previous section $\intd^{(8,2)}(q)$ comes from
the 4-loop integrand expressed in terms of $u$, when we take into account the
$g^2$ shifts of the rapidities of the constituents of the Konishi due to ABA. 
In order to find it we have to plug $u=\frac{1}{2\sqrt{3}}+\frac{4}{\sqrt{3}}
g^2$ into $\intd^{(8)}(q)$ and expand it to the second order in $g^2$. The result
is
\begin{eqnarray*}
&&\intd^{(8,2)}(q)=-\frac{4718592 Q^2  (16 + 9 q^4 - 12 Q^2 + 9 Q^4 + 
    6 q^2 (-10 + 3 Q^2)) }{(9 q^4 + 
    6 q^2 (2 - 6 Q + 3 Q^2) + (4 - 6 Q + 3 Q^2)^2)^2 } \times\\
    &&\hspace{1cm}\times \frac{(-4 + 3 q^2 + 3 Q^2)(-16 - 9 q^4 - 12 Q^2 + 27 Q^4 + 
    6 q^2 (-2 + 3 Q^2))}{(9 q^4 + 
    6 q^2 (2 + 6 Q + 3 Q^2) + (4 + 6 Q + 3 Q^2)^2)^2(q^2 + Q^2)^4 }
\end{eqnarray*}

\subsection{Integration}

Before we proceed, it is fruitful to observe that we can symmetrize the integrand
with
respect to $q$ without changing the result of integration\begin{equation}
\intd^{sym}(q)=\frac{1}{2}(\intd(q)+\intd(-q))\end{equation}
When we calculated the integrals we
symmetrized some parts of the integrand leaving the rest not symmetrized
depending on which form is easier to handle and gives simpler result.

Apart from the polygamma functions which appear in the dressing and scalar part,
the remaining part of the integrand
is a rational function. It can be then integrated over the real line by
taking residues at the position of poles lying above the real
line. All such poles can be classified into two groups: poles coming
from the S-matrix parts of the integrand ('dynamical'
poles), four of which lie above the real
line: 
\eq
q= i(Q\pm 1) \pm \frac{1}{\sqrt{3}} \ ,
\eqx
and a pole coming from the
exponential part which is exactly
$q=iQ$. 

It turns out that when we symmetrize the whole integrand the contribution coming
from dynamical poles vanishes when summed over $Q$ as in the 4-loop
case\footnote{It can be checked both numerically and completely algebraically
using the methods of the subsequent section.}. It can be explained by the fact
that
at weak coupling we expect the $\mu$-term contributions to be absent. We want to
stress that this is a far from trivial consistency check of our formulas.

Beside the residues of the rational function poles we
have to
handle the additional poles coming from the polygamma functions which appear in
the dressing and scalar
part. In order to simplify our considerations, it is convenient to
symmetrize the
$S_{0\psi}^{(2)}(q,Q,u)$ part. It turns out that in that case the polygamma
functions will vanish leaving us only with rational function
\begin{equation*}
\Big( \psi\big( \frac{1}{2}(-i q - Q)\big) - 
  \psi\big(\frac{1}{2} (-i q + Q)\big) \Big)^{sym}=-\frac{2 Q}{q^2 + Q^2}
\end{equation*}  
So the only remaining polygamma functions come from the dressing part. The
positions of
polygamma function poles are known to be at negative integers. In our
case it means that $\frac{1}{2}(-iq-Q)=-n$ or $\frac{1}{2}(iq+Q)=-n$ where
$n\geq 0$. Solving with respect to $q$ we obtain the positions of integrand
poles coming from polygamma function as
\begin{eqnarray}\label{psipoles}
q=i(Q-2n)\ \ \ n\geq 0\ \ \
\mbox{    or    }\ \ \
q=i(Q+2n)\ \ \ n\geq 0
\end{eqnarray} 
We want to close the contour of integration over the real line meaning we have
to take into account only those residues which have nonnegative imaginary part.

In the proceeding sections we will calculate all the residues. We have chosen to
symmetrize only two parts of integrand: the scalar integrand part which contains
polygamma functions and the part coming from the ABA modification. The rest
is left in a nonsymmetrized form. 

\subsubsection*{Residues of rational functions}

For the rational part of the integrand the result after taking the residues
(regardless if we take $q=iQ$ or 'dynamical' poles) can be rewritten as a sum of
terms
with minimal denominators which are of the form
\begin{equation}\label{sumtype1}
\frac{a}{Q^n}
\end{equation}  
which will give zeta functions when summed up or
\begin{equation}\label{sumtype2}
\frac{aQ+b}{1\pm 3Q+3Q^2}
\end{equation}
which will produce polygamma functions.

For the dressing part the residue at $q=iQ$ gives exactly
\begin{equation}
\frac{3456 (-2 + 12 Q^2 - 45 Q^4 + 27 Q^6) \zeta(3)}{Q^3 (1 - 3 Q^2 + 9 Q^4)^2}
\end{equation}
while the residues at the dynamical poles are some rational functions of $Q$
with complicated coefficients containing $\psi(\frac{1}{6} (3 + i \sqrt{3}))$
and $\psi(\frac{1}{6} (3 - i \sqrt{3}))$. 

\subsubsection*{Residues of polygamma functions}

To calculate the contribution coming from the polygamma functions appearing in
the dressing part of the integrand we want to find all the poles of the form
(\ref{psipoles}) which have nonnegative imaginary part. It is easy to notice
that such poles are of the form $q_{Q,n}=i(Q+2n)$  where $n$ is (possibly
negative) integer obeying $Q+2n\geq 0$. The residues at the positions $q_{Q,n}$
is found to be
\begin{eqnarray*}
\frac{-864 Q^2 (1 + 3 n^2 + 3 n Q)^2 \mbox{sign}(n)}{
 n^4 (1 - 3 n^2 + 9 n^4) (n + Q)^4 \big( (1 + 3 n^2 + 6 n Q + 3 Q^2)^2 - (3 n +
3 Q)^2\big)}
\end{eqnarray*}
The remaining task is to sum the above formula over $n$ and $Q$. We have to be
careful during the summation process because there exist one pole which lies on
the real line for every even $Q$. In that case we have to take only half of the
residue with the minus sign due to how we have chosen the orientation of our
integration contour. Additionally, we have to remember that the residues at
$q=iQ$ were taken into account before. Keeping it in mind we calculated the sum
over $Q$ and obtained
\begin{eqnarray*}
\frac{1728 ( 3 n^2-1) (-1 + 3 n^2 + 3 n^3 - 9 n^4 + 
   9 n^5 + (n^3 - 3 n^5 + 9 n^7)\psi^{(2)}( 1 + n))}{n^6 (1 - 3 n^2 + 9 n^4)^2}
\end{eqnarray*}
which have to be summed over $n$ from $1$ to $\infty$. It can be done using
methods from the next section.

\section{Summation over bound-states}

There are two types of sums over bound-states appearing in our calculations
after the residues are taken. Firstly, there are sums of the form
(\ref{sumtype1}) or (\ref{sumtype2}) which can be easily summed using
{\it Mathematica}. The first ones give us $\zeta$-functions while the second
ones contain polygamma functions of the form
$\psi(\frac{1}{6} (3 + i \sqrt{3}))$ and $\psi(\frac{1}{6} (3 - i \sqrt{3}))$.

On the other hand during the 5-loop Konishi calculation more difficult sums
emerge

\begin{equation}
\Sigma^{(m)}=\sum_{Q=1}^{\infty} \, R(Q) \, \psi^{(m)}(Q) \qquad m\geq 0,
\label{spgv1}
\end{equation}
where $R(x)$ is a rational function of $x$ and $\psi^{(m)}(x)$ is the $m$th
polygamma
function given by the definition
\begin{equation}
\psi^{(m)}(x)=\frac{d^m \,\psi(x)}{d \,x^m}, \qquad m\geq 1,
\end{equation}
with $\psi(x)\equiv\psi^{(0)}(x)$ being the digamma function $\psi(x)= \frac{d
\,\log \Gamma(x)}{d \,x}$.
The evaluation of the sum (\ref{spgv1}) goes as follows:
$R(Q)$ is decomposed as a sum of two terms $R(Q)=R_0(Q)+R_1(Q)$, where $R_0(Q)$
contains
the sum of pure power terms of the partial fraction decomposition of $R(Q)$
(i.e. $R_0(Q)=\frac{a_1}{Q^{n_1}}+\frac{a_2}{Q^{n_2}}+...$), while $R_1(Q)$
contains the rest.
 In this case the sum (\ref{spgv1}) is decomposed into two parts as well:
 \begin{equation}
\Sigma^{(m)}=\Sigma_0^{(m)}+\Sigma_1^{(m)},
\end{equation}
where
\begin{equation}
\Sigma_a^{(m)}=\sum_{Q=1}^{\infty} \, R_a(Q) \, \psi^{(m)}(Q) \qquad m\geq 0,
\qquad a=0,1.
\label{spga}
\end{equation}
Using a series representation for the polygamma functions,
$\Sigma_0^{(m)}$ can be expressed in terms of the values at infinity of nested
harmonic sums,
which can be expressed in terms of multivariate zeta functions (Zagier-Euler
sums) \cite{KV05}.
These sums can be reexpressed in terms of ordinary Euler sums. The relations
between the
former and the latter can be found using EZ-Face - an online calculator for Euler
sums \cite{EZFace}.
At the end of the process $\Sigma_0^{(m)}$ is expressed in terms of zeta
functions taken at integer
values.

 The calculation of $\Sigma_1^{(m)}$ requires a different method:
representing $R_1(x)$ as the Laplace transform of its inverse Laplace transform,
and using
appropriate integral representation for the polygamma functions,
$\Sigma_1^{(m)}$ can be transformed into a double integral form, in which the
summation can
be easily performed and the remaining double integral expression can be
calculated exactly
with the help of {\it Mathematica}. For details see Appendix B.

Using the methods presented above all the sums can be evaluated. The striking
observation is that all `nasty' polygamma functions disappear leaving us only
with the $\zeta$-functions. The final result is

\begin{equation}
\Delta_w^{(10)} =-11340+2592\zeta(3) -5184\zeta(3)^2 -11520\zeta(5)
+30240\zeta(7)
 \end{equation}  

As for the 4-loop case the result is a sum of $\zeta$-functions of odd
degrees with integer coefficients and correct transcendentality degree. The new
feature of the result is that products of $\zeta$-functions start to appear
which have been absent in the 4-loop case.   

To summarize the anomalous dimension of the Konishi operator up to 5-loops is
\eqn
\Delta&=&4+12g^2-48g^4+336 g^6+ 96 (-26 + 6 \zeta(3) - 15 \zeta(5)) g^8 
 \\ \nonumber
&& -96 (-158 - 72 \zeta(3) + 54 \zeta(3)^2 + 90 \zeta(5) - 315 \zeta(7)) g^{10}
\eqnx

\section{Conclusions}

Anomalous dimensions of operators in $\nn=4$ SYM correspond to energies of string states in $AdS_5 \times S^5$. The leading perturbative orders are given by the Asymptotic Bethe Ansatz, while at a certain loop order, there appear new contributions coming from a topologically distinct class of Feynman diagrams -- so-called `wrapping interactions'. On the string theory side these correspond to virtual particles propagating around the string worldsheet cylinder. The leading corrections arise from multiparticle generalizations of the classical L{\"u}scher terms. Further corrections are due to many virtual particles and all these should in principle be resummed by a TBA system.

The most convenient testing ground for these issues is the shortest nonprotected operator in $\nn=4$ SYM -- the Konishi operator, as well as single impurity operators in the $\beta$-deformed theory.

In a previous paper \cite{Bajnok:2008bm}, the leading four loop part of the wrapping correction has been found. It came from a two-particle L{\"u}scher term and gave
\eq
\Delta_{w,Konishi}^{(8)} = 324 +864\, \zeta(3) -1440 \,\zeta(5)
\eqx
The aim of the present paper was to compute the five loop wrapping contribution. Although it still arises just from the two-particle L{\"u}scher term, it is very interesting as it involves two new ingredients. Firstly, there is a nontrivial modification of the Asymptotic Bethe Ansatz quantization due to the sea of virtual particles. This term is quite interesting as it does not seem to be simply related to the transfer matrix appearing in a direct expansion of the proposed TBA systems. Secondly, the BES/BHL dressing factor cannot be neglected any more, and moreover, due to the specific kinematics of the L{\"u}scher term, an \emph{infinite} set of coefficients of the dressing phase starts to contribute at once. The result obtained in the present paper is
\eq
\Delta_{w,Konishi}^{(10)} =-11340+2592\,\zeta(3) -5184\,\zeta(3)^2 -11520\,
\zeta(5)
+30240\,\zeta(7)
\eqx
This computation is subject to two nontrivial cross-checks. Firstly, the partial results (coming from the dressing phase, modification of the ABA and the remaining part of the S-matrix) have a very complicated transcendentality structure involving polygamma functions. All these cancel out leaving the final result as a simple combination of odd $\zeta$ functions. Secondly, the residues of the `dynamical' poles (associated with $\mu$-terms) should cancel out between the various terms when summed over the types of bound-states. Both of these cross-checks are satisfied in our case and involve a rather intricate conspiracy between the various terms.

In Appendix A, we have computed the five loop wrapping correction to a single impurity in the $\beta$-deformed theory\footnote{This computation relies on an additional assumption on performing analytical continuation from even to odd spins in \cite{Bajnok:2008qj}. The four loop result has been verified by a direct perturbative computation in \cite{Fiamberti:2008sn}.}. In this case the four loop result was ($M=1$ in \cite{Bajnok:2008qj}, see also \cite{Gunnesson:2009nn,Beccaria})
\eq
\Delta_{w,single}^{(8)}=496\,\zeta(3)-640\,\zeta(5)
\eqx
while the five loop wrapping contribution computed in Appendix~A is
\eq
\Delta_{w,single}^{(10)}=-1536\,\zeta(3)^{2}-4096\,\zeta(3)-5120\,\zeta(5)+13440
\,\zeta(7)
\eqx
The ABA modification does not appear here, however the \emph{infinite} set of BES coefficients contribute just as in the case of the Konishi operator.

It would be very interesting to verify these results perturbatively. Especially the single impurity operator may be within reach. This might help in understanding the precise relation between the structure of direct perturbative computations and the string theory computations based on the S-matrices and 
L{\"u}scher formulas.

Another application of these results would be to test the excited state TBA systems recently proposed for AdS/CFT. It would be very interesting to see the appearance of the rather intricate modification of the Asymptotic Bethe Ansatz and the specific analytical continuation of the dressing phase from these formulations, especially as both of these ingredients are very strongly constrained by the transcendentality structure and the cancellation of dynamical poles between themselves and the rest of the integrand. The four loop result is sensitive to the TBA source terms and has been rederived in \cite{Gromov:2009bc}. The five loop result depends on the structure of the convolution terms and thus is a more sensitive test of excited state TBA systems. In particular the procedure of section 3 to obtain the L\"uscher corrections could be applied e.g. to the TBA system in the second paper of \cite{Gromov:2009bc}. However this is technically quite involved. The five loop wrapping correction for the Konishi operator seems therefore to be an interesting and robust test for the excited TBA systems.  

\bigskip

\noindent{\bf Acknowledgments:} We would like to thank Sergey Frolov, Christoph Sieg and Dima Volin for interesting discussions related to this project. R.J. and T.L. were supported by Polish science funds during 2009-2011 as a research project (NN202 105136). The work was supported by Marie Curie Transfer of Knowledge Project COCOS (MTKD-CT-2004-517186). R.J. was supported by the RTN network ENRAGE MRTN-CT-2004-005616.
Z.B. was supported by a Bolyai Research Fellowship and by the Hungarian National Science Fund OTKA T60040  and A.H. by OTKA T049495. 

\pagebreak

\appendix

\section{L\"uscher correction for a one particle state }

In this Appendix we calculate the subleading wrapping correction to
the energy of a one particle state. Such a state must have vanishing
momentum in the $\mathcal{N}=4$ super-Yang-Mills theory thus its
energy is protected. In the $\beta$ deformed theory, however, at
$\beta=\frac{1}{2}$ a one particle state with vanishing rapidity
$u=0$ ($p=\pi$) is allowed and acquires nontrivial finite size corrections.
The condition for its vanishing rapidity is protected at all orders
both in the ABA and in the wrapping corrections. 
So in this case we will not have a contribution coming from the modification of
the ABA quantization. However there will be a nontrivial
direct wrapping correction to the energy, which will include, starting from
five loops, a contribution from the BES dressing phase. Thus the
single
impurity operator provides a testing ground for that part of the L\"uscher
correction.

The leading order wrapping correction can be written as \eq
\Delta
E_{w}=-\sum_{Q=1}^{\infty}\int\frac{dq}{2\pi}(-1)^{F}S_{Q1}^{Q1}(q,0)e^{-\tilde{
\epsilon}(q)L}=-\sum_{Q=1}^{\infty}\int\frac{dq}{2\pi}Y_{Q}(q)\eqx
Here $S_{Q1}^{Q1}$ represents the scattering matrix of the mirror
$Q$ particle with the fundamental $u=0$ physical particle. We can
further decompose the scattering part as the scalar part ($S_{0}$),
the dressing part ($S_{\sigma}$) and the matrix part ($S_{\boxplus}$):
\eq
S_{Q1}^{Q1}(q,0)=S_{0}(q,Q)S_{\sigma}(q,Q)S_{\boxplus}(q,Q)^{2}\eqx
We expand each quantity to subleading order in $g^{2}$ as\begin{eqnarray*}
S_{0}(q,Q) & = & S_{0}^{(0)}(q,Q)+g^{2}S_{0}^{(2)}(q,Q)+\dots\\
S_{\sigma}(q,Q) & = & 1+g^{2}S_{\sigma}^{(2)}(q,Q)+\dots\\
S_{\boxplus}(q,Q) & = &
g^{2}S_{\boxplus}^{(2)}(q,Q)+g^{4}S_{\boxplus}^{(4)}(q,Q)+\dots\end{eqnarray*}
When we calculated the leading wrapping correction to twist two operators
for odd particle number \cite{Bajnok:2008qj}, we observed that in order to be compatible
with gauge theory calculations and to provide the proper analytical
continuation from even cases to odd ones we had to omit the contribution
of the fermions. We accept this convention now too, but call the attention
for the need of a derivation of this proposal based on first principles.
Under this assumption the leading order correction of the matrix part
is \eq
S_{\boxplus}^{(2)}(q,Q)=-\frac{16Q(q^{2}+Q^{2}-1)}{(q^{2}+Q^{2})(Q+iq-1)}\eqx
while the subleading one is 
\eq
\frac{S_{\boxplus}^{(4)}(q,Q)}{S_{\boxplus}^{(2)}(q,Q) }
=-\frac{4(q^{3}-iq^{2}(Q-3)-i(Q-1)^{3}+q(Q^{2}+1))}{(q^{2}
+Q^{2})(q-i(Q-1))}\eqx
The leading and subleading correction of the exponential part is given
by \eq
e^{-\tilde{\epsilon}(q)}=g^4 \ex ^{(4)}(q,Q)+g^6 \ex ^{(6)}(q,Q)
=\frac{16g^4}{(q^{2}+Q^{2})^{2}}\left[1-g^{2}\frac{16}{(q^{
2}+Q^{2})}\right]\eqx
In calculating the scalar part we have to use the parameterization
of the bound-state as we did for the Konishi operator. The result for
the leading order scalar part reads as \eq
S_{0}^{(0)}(q,Q)=-\frac{q-i(Q-1)}{(q+i(Q-1))(q-i(Q+1))(q+i(Q+1))}\eqx
while the subleading contains a rational part\begin{eqnarray*}
\frac{S_{0rat}^{(2)}(q,Q)}{S_{0}^{(0)}(q,Q)} & = &
8+\frac{16 (-1 + 2 i q) q^2 - 
 16 (2 + q (i + 6 q)) Q}{(1 + 4 q^2) (q^2 + Q^2)}\\ & &
+ \frac{16q i (-1 + 2 q^2 + Q)}{(1 + 4 q^2) (q^2 + (-1 + Q)^2)}-\frac{2 q (3 + 2
Q)}{(1 + 4 q^2) (q^2 + (1 + Q)^2))}
\end{eqnarray*}
and a polygamma part: \eq
\frac{S_{0\psi}^{(2)}(q,Q)}{S_{0}^{(0)}(q,Q)}
=8\psi(-\frac{1}{2}(iq+Q))-8\psi(\frac{1}{2}(-iq+Q))\eqx
The dressing part reads as \eq
S_{\sigma}(q,Q)=-8\left[2\gamma_{E}+\psi(-\frac{1}{2}(iq+Q))+\psi(\frac{1}{2}
(iq+Q))\right]
\eqx
The full integrand can be written as
\eq
Y_Q(g)=g^8 Y_Q^{(8)}(q)+g^{10} Y_Q^{(10)}(q)+\dots 
\eqx
The leading order part is given by \begin{eqnarray*}
Y_{Q}^{(8)}(q) & = &
S_{0}^{(0)}(q,Q)S_{\boxplus}^{(2)}(q,Q)^{2}\Upsilon^{(4)}(q,Q)\\
 & = &
-\frac{4096Q^{2}(-1+q^{2}+Q^{2})^{2}}{(q^{2}+Q^{2})^{4}(q^{4}+(-1+Q^{2})^{2}+2q^
{2}(1+Q^{2}))}\end{eqnarray*}
It has a kinematical pole at $iQ$ and two dynamical poles at $i(Q\pm1)$
on the upper half plane.
If we take the residue of the integrand only at the kinematical pole
and sum over $Q$ we obtain the leading wrapping correction 
\eq
\Delta_{w}^{(8)}=128(4\zeta(3)-5\zeta(5))
\eqx
We note that the contributions of the dynamical poles summed
over $Q$ cancel out, so this is indeed the full leading order result, which
has been verified by a direct perturbative calculation on the gauge theory side in \cite{Fiamberti:2008sn}.

In calculating the 5-loop subleading correction we can write\eqs 
Y_{Q}^{(10)}(q)=Y_{Q}^{(8)}(q)\left[\frac{S_{0}^{(2)}(q,Q)}{S_{0}^{(0)}(q,Q)}
+2\frac
{S_{\boxplus}^{(4)}(q,Q)}{S_{\boxplus}^{(2)
}(q,Q)}+S_{\sigma}^{(2)}(q,Q)+\frac{\Upsilon^{(6)}(q,Q)}{\Upsilon^{(4)}(q,Q)}
\right]\eqsx
We analyze separately the contributions of the rational part and the
polygamma part. We can observe that the rational part of the scalar
contribution together with the matrix contribution and the exponential
part gives a symmetric function in $q$: \begin{eqnarray*}
\frac{S_{0rat}^{(2)}(q,Q)}{S_{0}^{(0)}(q,Q)}+2\frac{S_{\boxplus}^{(4)}(q,Q)}{S_{
\boxplus}^{(2)}(q,Q)}+\frac{\Upsilon^{(6)}(q,Q)}{
\Upsilon^{(4)}(q,Q)} & =\\
 &  &
\hspace{-6cm}-\frac{(8(q^{4}(7+2Q)+(3+2Q)(-1+Q^{2})^{2}+2q^{2}(5+2Q)(1+Q^{2}))}{
(q^{2}+(-1+Q)^{2})(q^{2}+Q^{2})(q^{
2}+(1+Q)^{2}))}\end{eqnarray*}
Taking the residue at the kinematical pole and summing over $Q$ gives
\eq
\frac{128}{27}(72\pi^{2}-30\pi^{4}+2\pi^{6}
-1296\zeta(3)-1080\zeta(5)+2835\zeta(7))
\eqx
The contributions of the dynamical poles do not cancel when summed
over $Q$ (as we did not analyze yet the full expression) and give 
\eq
-\frac{256}{3}(-12\pi^{2}+\pi^{4})
\eqx
The polygamma part is a bit more complicated as it has poles not only
at the kinematical and dynamical locations but additionally at\footnote{In contrast to
the Konishi case we have not symmetrized the polygamma part of scalar
integral. This choice leads to a slightly different configuration of the poles.} $q=i(Q+2n)$,
$n>0$.
The contribution of its residues are \begin{eqnarray*}
\mathop{\mbox{Res}}_{q=i(Q+2n)}Y_{Q}^{(8)}(q)\left[\frac{S_{0\psi}^{(2)}(q,Q)}{
S_{0}^{(0)}
(q,Q)}+S_{\sigma}^{(2)}(q,Q)\right] & =\\
 &  &
\hspace{-3cm}-\sum_{n=1}^{\infty}\frac{256Q^{2}(1+4n(n+Q))^{2}}{n^{4}(-1+4n^{2}
)(n+Q)^{4}(-1+4(n+Q)^{2}))}\end{eqnarray*}
This can be summed over $n$ resulting in polygamma functions. We
combine this result with the contribution of the remaining poles obtained by
taking residues of
$Y_{Q}^{(8)}(q)\left[\frac{S_{0\psi}^{(2)}(q,Q)}{S_{0}^{(0)}(q,Q)}+S_{\sigma}^{
(2)}
(q,Q)\right]$ at
the kinematical and dynamical poles. The final expression can be summed
over $Q$ by the methods of Appendix B. The result is simply
\eq
\frac{256}{27}(-\pi^{2}(-12+\pi^{2})^{2}+54(4-3\zeta(3))\zeta(3))\eqx
 When we combine this result with the result of the rational part
we can observe that the even $\zeta$ part cancels out and we arrive at
the subleading wrapping correction
\eq
\Delta_{w}^{(10)}=-128(12\zeta(3)^{2}+32\zeta(3)+40\zeta(5)-105\zeta(7))
\eqx
We checked that the dynamical residue contributions of
the full integrand cancel when summed over $Q$ as in the Konishi
case providing another convincing support for our considerations.

\section{Summation of terms containing polygamma functions}

Here we present a method which enables one to calculate sums containing
polygamma functions.
A typical sum emerging during the 5-loop Konishi computation is of the form:
\begin{equation}
\Sigma^{(m)}=\sum_{Q=1}^{\infty} \, R(Q) \, \psi^{(m)}(Q) \qquad m\geq 0,
\label{spgB}
\end{equation}
where $R(x)$ is a rational function of $x$ and $\psi^{(m)}(x)$ is the $m$th
polygamma
function given by the definition
\begin{equation}
\psi^{(m)}(x)=\frac{d^m \,\psi(x)}{d \,x^m}, \qquad m\geq 1,
\end{equation}
with $\psi(x)\equiv\psi^{(0)}(x)$ being the digamma function $\psi(x)= \frac{d
\,\log \Gamma(x)}{d \,x}$.
The evaluation of sum (\ref{spgB}) goes as follows:
$R(Q)$ is decomposed as a sum of two terms $R(Q)=R_0(Q)+R_1(Q)$, where $R_0(Q)$
contains
the sum of pure inverse power terms of the partial fraction decomposition of $R(Q)$
(i.e. $R_0(Q)=\frac{a_1}{Q^{n_1}}+\frac{a_2}{Q^{n_2}}+...$), while $R_1(Q)$
contains the rest.
 In this case the sum (\ref{spgB}) is decomposed into 2 parts as well:
 \begin{equation}
\Sigma^{(m)}=\Sigma_0^{(m)}+\Sigma_1^{(m)},
\end{equation}
where
\begin{equation}
\Sigma_a^{(m)}=\sum_{Q=1}^{\infty} \, R_a(Q) \, \psi^{(m)}(Q) \qquad m\geq 0,
\qquad a=0,1.
\label{spgaB}
\end{equation}
Using a series representation for the polygamma functions the sum
$\Sigma_0^{(m)}$ can be evaluated
directly applying the method sketched in section 6. So, hereafter we concentrate
on the calculation
of the sums $\Sigma_1^{(m)}$. They are evaluated by transforming them into
integral
expressions calculable
with the help of {\it Mathematica}.
The transformation is as follows:
the polygamma functions are represented by their appropriate integral
representations
\begin{equation} \label{psim}
\psi^{(m)}(x)=(-1)^{m+1} \, \int\limits_0^{\infty} dt \, \frac{t^m \,
e^{-xt}}{1-e^{-t}}, \qquad m\geq 1,
\end{equation}
\begin{equation} \label{psi0}
\psi(x)\equiv\psi^{(0)}(x)= \int\limits_0^{\infty} dt \,
\left(\frac{ e^{-t}}{t}- \frac{ e^{-xt}}{1-e^{-t}} \right).
\end{equation}
and the function $R_1(x)$ is represented as the Laplace transform of its inverse
Laplace transform
\begin{equation} \label{ILR1}
R_1(x)=\int\limits_0^{\infty} dt \, e^{-xt} \, {\cal L}R_1^{-1}(t),
\end{equation}
where ${\cal L}R_1^{-1}(t)$ stands for the inverse Laplace transform of $R_1(x)$
given by the formula
\begin{equation}
{\cal L}R_1^{-1}(t)=\frac{1}{2\pi i} \int\limits_{\eta-i \infty}^{\eta+i \infty}
ds \,
e^{st} \, R_1(s),
\end{equation}
with $\eta$ being an arbitrary positive constant chosen so that the contour of
integration lies
to the right of all singularities in $R_1(s)$. Due to the structural
difference between
integral representations (\ref{psim}) and (\ref{psi0}) one has to make
a distinction between the
cases $m\geq1$ and $m=0$ and consider them separately. \newline
{\bf \underline{$m\geq 1$ case:} } \newline
Using integral representations (\ref{psim}) and (\ref{ILR1}), $\Sigma_1^{(m)}$
takes the form
\begin{equation}
\Sigma_1^{(m)}=\sum_{Q=1}^{\infty} \int\limits_0^{\infty} dt \, e^{-Qt} \, {\cal
L}R_1^{-1}(t) \,
(-1)^{m+1} \, \int\limits_0^{\infty} dt' \, \frac{t'^m \, e^{-Qt'}}{1-e^{-t'}}
\nonumber
\end{equation}
which after evaluating the simple geometric sum in $Q$ becomes
\begin{equation}
\int\limits_0^{\infty} dt \, {\cal L}R_1^{-1}(t) \,(-1)^{m+1}
\,\int\limits_0^{\infty} dt' \,
 \frac{t'^m \, }{(1-e^{-t'})(e^{t+t'}-1)}
\nonumber
\end{equation}
Then using the identity
\begin{equation}
\int\limits_0^{\infty} dt' \, \frac{t'^m \,
}{(1-e^{-t'})(e^{t+t'}-1)}=
-\Gamma(m+1) \, \frac{\mbox{Li}_{m+1}(e^{-t})-\zeta(1+m)}{e^t-1},
\end{equation}
where $\mbox{Li}_n(x)$ is the $n$th polylogarithm function,
the integral with respect to $t'$ can be evaluated and finally a single integral
remains
\begin{equation}  \label{S1m}
\Sigma_1^{(m)}=(-1)^m \, \int\limits_0^{\infty} dt \, {\cal L}R_1^{-1}(t) \,
\Gamma(m+1)
\, 
\frac{\mbox{Li}_{m+1}(e^{-t})-\zeta(1+m)}{e^t-1}.
\end{equation}
During the Konishi computation $m$ took the values of $1,2$ and $3$ and in
all cases emerging
during the calculations, the integrals (\ref{S1m}) could be evaluated by
{\it Mathematica}.
\newline
{\bf \underline{$m=0$ case} } \newline
Representing $R_1(x)$ as the Laplace transform of its inverse Laplace transform,
and using integral
representation (\ref{psi0}) for $\psi^{(0)}(x)$ the sum takes the form:
\begin{equation}
\Sigma_1^{(0)}=\sum_{Q=1}^{\infty} \int\limits_0^{\infty} dt \, e^{-Qt} \, {\cal
L}R_1^{-1}(t) \,
\int\limits_0^{\infty} dt' \,
\left(\frac{ e^{-t'}}{t'}- \frac{ e^{-Qt'}}{1-e^{-t'}} \right)
\nonumber
\end{equation}
Again evaluating the simple geometric sum in $Q$, this can be recast as
\begin{equation}
\int\limits_0^{\infty} dt  \, {\cal L}R_1^{-1}(t) \,
\int\limits_0^{\infty} dt' \,
\left(\frac{ e^{-t'}}{t' \, (e^t-1)}- \frac{1}{(1-e^{-t'})\,(e^{t+t'}-1)}
\right) \nonumber
\end{equation}
Now exploiting the integral formula
\begin{equation}
\int\limits_0^{\infty} dt' \,
\left(\frac{ e^{-t'}}{t' \, (e^t-1)}- \frac{1}{(1-e^{-t'})\,(e^{t+t'}-1)}
\right)=
\frac{\gamma_E + \log(1-e^{-t})}{1-e^{t}},
\end{equation}
with $\gamma_E$ being the Euler's constant,
the integral with respect to $t'$ can be evaluated  and one ends up with an
expression
containing a single integral
\begin{equation} \label{S0}
\Sigma_1^{(0)}=\int\limits_0^{\infty} dt  \, {\cal L}R_1^{-1}(t) \,
\frac{\gamma_E + \log(1-e^{-t})}{1-e^{t}}.
\end{equation}
During Konishi computations this formula made it possible to evaluate sums
containing $\psi^{(0)}(x)$ 
functions.

Finally, we just note that in the case of $m=0$ the sum $\Sigma_0^{(0)}$ can
also be
evaluated by the application
of formula (\ref{S0}).

\pagebreak

\end{document}